This work has been extended and is now published in Journal of Applied Clinical Medical Physics (JACMP, 2025-08-24, http://dx.doi.org/10.1002/acm2.70221)

# Impact of deep learning model uncertainty on manual corrections to auto-segmentation in prostate cancer radiotherapy

Viktor Rogowski[a,f], Angelica Svalkvist[g,h], Matteo Maspero[c,d], Tomas Janssen[e], Federica Carmen Maruccio[e], Jenny Gorgisyan[a,f], Jonas Scherman[a], Ida Häggström[j,g], Victor Wåhlstrand[j], Adalsteinn Gunnlaugsson[i], Martin P Nilsson[i], Mathieu Moreau[i], Nándor Vass[i], Niclas Pettersson[g,h], Christian Jamtheim Gustafsson[a,b*]

[a] Radiation Physics, Department of Hematology, Oncology, and Radiation Physics, Skåne University Hospital, Klinikgatan 5, Lund 221 85, Sweden
[b] Department of Translational Medicine, Medical Radiation Physics, Lund University, Carl Bertil Laurells gata 9, Malmö 205 02, Sweden
[c] Department of Radiation Oncology, Imaging and Cancer Division, University Medical Center Utrecht Utrecht, The Netherlands
[d] Computational Imaging Group for MR Diagnostics & Therapy, Center for Image Sciences University Medical Center Utrecht, Utrecht, The Netherlands
[e] Department of Radiation Oncology, The Netherlands Cancer Institute Antoni van Leeuwenhoek, Amsterdam, The Netherlands
[f] Department of Medical Radiation Physics, Clinical Sciences, Lund University, Lund, Sweden
[g] Department of Medical Radiation Sciences, Institute of Clinical Sciences, Sahlgrenska Academy, University of Gothenburg, Gothenburg, Sweden
[h] Region Västra Götaland, Sahlgrenska University Hospital, Department of Medical Physics and Biomedical Engineering, Gothenburg, Sweden
[i] Department of Hematology, Oncology, and Radiation Physics, Skåne University Hospital, Lund, Sweden
[j] Department of Electrical Engineering, Chalmers University of Technology, Gothenburg, Sweden

* Corresponding author: christian.jamtheimgustafsson@skane.se
Radiation Physics, Department of Hematology, Oncology, and Radiation Physics
Skåne University Hospital
Klinikgatan 5
Lund 221 85, Sweden

**Abstract**

**Background:** Deep learning (DL)-based organ segmentation is increasingly used in radiotherapy. While methods exist to generate voxel-wise uncertainty maps from DL-based auto-segmentation models, these maps are rarely presented to clinicians.

**Purpose:** This study aimed to evaluate the impact of DL-generated uncertainty maps on experienced radiation oncologists during the manual correction of DL-based auto-segmentation for prostate radiotherapy.

**Methods:** Two nnUNet DL models were trained with 10-fold cross-validation on a dataset of 434 patient cases undergoing ultra-hypofractionated MRI-only radiotherapy for prostate cancer. The models performed prostate clinical target volume (CTV) and rectum segmentation. Each cross-validation model was evaluated on an independent test set of 35 patient cases. Segmentation uncertainty was calculated voxel-wise as the SoftMax standard deviation (0-0.5, n=10) and visualized as a fixed scale color-coded map.

Four experienced oncologists were asked to:

Step 1: Rate the quality of and confidence in the DL segmentations using a four- and five-point Likert scale, respectively, and edit the segmentations without access to the uncertainty map.

Step 2: Repeat step 1 after at least four weeks, but this time with the color-coded uncertainty map available.

Oncologists were asked to blend the uncertainty map with the DL segmentation and MRI volume. Segmentation edit time was recorded for both steps. In step 2, oncologists also provided free-text feedback on the benefits and drawbacks of using the uncertainty map during segmentation. A histogram analysis was performed to compare the number of voxels edited between step 1 and step 2 for different uncertainty levels (bins with 0.1 intervals).

**Results:** The DL models achieved high-quality segmentations with a mean Dice coefficient per oncologist of 0.97-0.99, calculated between edited and unedited segmentation in step 1 for the prostate CTV and rectum.

While the overall quality rating for rectum segmentations decreased slightly on a group level in step 2 compared to step 1, individual responses varied. Some oncologists rated the quality higher for the prostate CTV segmentation with the uncertainty map present, while others rated it lower. Similarly, confidence ratings varied across oncologists for prostate CTV and rectum.

Decreased segmentation time was recorded for three oncologists using uncertainty maps, saving 1-2 minutes per patient case, corresponding to 14%-33% time reduction. Three oncologists found the uncertainty maps helpful, and one reported benefit was the ability to identify regions of interest more quickly. The histogram analysis had fewer voxel edits in regions of low uncertainty in step 2 compared to step 1. Specifically, 50% fewer voxel edits were recorded for the uncertainty region 0.0-0.1, suggesting increased trust in the DL model's prediction in these areas.

**Conclusions:** Presenting DL uncertainty information to experienced radiation oncologists influences their decision-making, quality perception, and confidence in the DL segmentations. Regions with low uncertainty were less likely to be edited, indicating increased reliance on the model's predictions. Additionally, uncertainty maps can improve efficiency by reducing segmentation time. DL-based segmentation uncertainty can be a valuable tool in clinical practice, enhancing the efficiency of radiotherapy planning.

## 1. Introduction

Integrating deep learning (DL)-based software in healthcare promises to substantially enhance patient care. By speeding up processes, ensuring accurate and individualized medical interventions, and providing a standardized interpretation of medical data, DL can decrease subjective differences among clinicians [1-3]. However, it is important that the decisions and uncertainties from the models can be understood and assessed, especially for healthcare applications where misdiagnoses or errors can be detrimental. The treatment planning process for radiation therapy, which contains several time-consuming and manual steps, can particularly benefit from DL-based solutions. For instance, manual segmentation of organs-at-risk (OARs) is both time-consuming and subject to inter-observer variability, as has been demonstrated, for example, for prostate [4], head-and-neck [5], and brain [6]. DL-based segmentation for radiotherapy planning reduces inter-observer variability [7] and decreases segmentation time [7-10]. This reduction in variability saves time and ensures a more consistent and reliable approach to the treatment planning process.

DL models in healthcare are often considered "black boxes," as their decision-making processes are opaque and difficult to understand. This lack of transparency raises trust and reliability issues in critical healthcare settings [11] and stresses the need to quantify and convey model uncertainty. For radiation therapy planning, several papers have been presented where segmentation uncertainty has been calculated and visualized for head-and-neck [12-14], prostate [15,16], and lung [17] . These methods are often based on Monte Carlo dropout [12-19], but recurring to deep ensembles have also been reported with a similar ability to predict uncertainty [13,19]. The benefit of deep ensembles is eliminating the need to add dropouts to the model training. The uncertainty probability is often calculated using entropy [12,13,17] or simply by using the SoftMax output of the models themselves [14-16]. While the latter approximation leads to less calibrated probabilities, its estimates are sufficient for many objectives [20] and are hence also used in this work.

Several papers have identified the need to assess the impact of segmentation uncertainty in the clinical setting [14,16,19], including its impact on segmentation time [17,21]. To the best of the authors' knowledge, no studies have examined the impact of presenting DL-based segmentation uncertainty on oncologists' behavior in a clinical setting. Consequently, the clinical implications of displaying segmentation uncertainty remain unknown.

In this study, we assess the impact of presenting DL-based segmentation uncertainty for the prostate CTV and rectum on experienced radiation oncologists in a clinical setting. The impact was quantitatively assessed on a voxel level with geometric metrics and measurement of segmentation time. This was accompanied by qualitative questions and free-text answers to investigate how the segmentation uncertainty affected the opinion and level of confidence in oncologist decision-making.

## 2. Methods

### 2.1 Data and model training

Separate DL models for segmentation of the prostate as a clinical target volume (CTV) and rectum as an OAR on T2-weighted (T2w) MR images were developed using the nnUNet [22] framework (version 2). The training and validation dataset consisted of 434 patient cases who received ultra-hypofractionated MRI-only based external radiotherapy [23] for intermediate and high-risk prostate cancer at a dose prescription of 42.7 Gy in 7 fractions, excluding the seminal vesicles, at Skåne University Hospital in Lund, Sweden. The selection criteria for extracting the data were as follows: no hip metal implants, the existence of exactly one MRI scan session, exactly one DICOM treatment plan, exactly one DICOM structure set for target and OAR, and exactly one structure set for fiducial markers. The MRI data was acquired on a 3T GE Healthcare MRI scanner and consisted of a large field of view (FOV) (44.8 cm-48 cm FOV) transversal T2w image encompassing the whole patient contour and was acquired with a 2D-based acquisition with an in-plane resolution of 0.6x0.7 to 0.8x0.9 mm and a slice thickness of 2.5 to 2.8 mm. Two versions of the acquisition protocols were used to collect 325 and 109 patient cases, respectively, with details provided in [23,24]. ESTRO guidelines [25] were used for the prostate CTV and rectum segmentations created during the clinical treatment planning process. Data was anonymized after extraction. A holdout test set of 35 consecutive patient cases was extracted with the same clinical inclusion criteria as the training/validation data and acquired with the acquisition protocol in Olsson, Af Wetterstedt, Scherman, Gunnlaugsson, Persson and Jamtheim Gustafsson [24]. The models were trained using 10-fold cross-validation for both organs. Source code for preprocessing and training can be found at https://github.com/jamtheim/segmentationUncertaintyPublic.

Ethics approval for this study was provided by the regional ethics board “Regionala Etikprövningsnämnden i Lund”, diary number 2013/742, and the Swedish Ethical Review Authority diary number 2024-01720-02. MR image data and segmentations have been publicly released in a dataset called LUND-PROBE and can be requested at

https://datahub.aida.scilifelab.se/10.23698/aida/lund-probe. The dataset is described in a separate pre-published paper [26].

## 2.2 Model inference and generation of the uncertainty map

During inference, nnUNet, by default, applies each model from each cross-fold validation on each patient case of the test dataset and then performs a model ensemble using average SoftMax across models for the final segmentation output (prostate CTV and rectum handled separately). To estimate the uncertainty for the prostate CTV and rectum model, the voxel-wise SoftMax values (0 to 1 range) from each cross-validation model inference were stored. The SoftMax sample standard deviation between the ten cross-validation models was then calculated in each voxel for the prostate CTV and rectum, respectively. It was desired that the uncertainty map was visualized in color and overlaid on the MR image using the same color scale for all patient cases in VARIAN Eclipse (v.15.6, Varian Medical Systems, Palo Alto, USA), thereby maintaining the software used in clinical routine (Fig.1).

To achieve the same color scale for all patient cases, a rectangular block with a value of 0.5, asserted to be the highest value for the voxel SoftMax standard deviation in the test data cohort (0 to 0.5 range), was inserted at the bottom of each image slice. The uncertainty map volume was saved as a positron emission tomography (PET) encoded DICOM image stack with a maintained frame of reference (Fig.1). This allowed for a simple overlay between a colorized uncertainty map and MRI volume.

## 2.3 Observer study

An observer study, consisting of two steps, was conducted where five experienced radiation oncologists from Skåne University Hospital were recruited to review and edit both the prostate CTV and the rectum segmentations originating from the DL nnUNet models from the 35 patient cases in the test set. One oncologist withdrew participation before the study commenced due to an inability to allocate sufficient time within his clinical schedule. The participating four oncologists had seven, nine, eleven, and eighteen years of segmentation experience, respectively. Segmentations from the nnUNet models will hereafter be referred to as DL segmentations, and the oncologists will be referred to as obsB, obsC, obsD, and obsE. The uncertainty map was not available to the oncologists in the first part of the study (Step 1). In the second part of the study (Step 2), the uncertainty map was overlaid on the MRI with the DL segmentations. Before the start of the study, a consensus meeting was held with the oncologists, who were instructed to segment according to clinical routine, i.e., as training data was segmented using ESTRO guidelines [25]. The oncologists were instructed not to discuss the patient cases with each other during the study. The written instructions provided to the oncologists during the study are attached in the supplementary material.

#### 2.3.1 Step 1 – Editing and assessment of DL segmentations without uncertainty map

The MRI with DL-generated segmentations for the prostate CTV and rectum was imported for all 35 test cases in Eclipse without any uncertainty map. A copy of this data was prepared for each oncologist, and the patient case order in the study was randomized for each oncologist. The oncologist was asked to assess the quality of the DL segmentations for the prostate CTV and rectum and edit the segmentations as needed. The oncologist recorded the time needed to edit both segmentations per patient case. For each segmentation and case, the following questions were answered on a Likert scale by the oncologist:

- Q1. How do you rate the unedited segmentation of the prostate CTV with respect to acceptance for treatment planning?
- Q2. How do you rate the unedited segmentation of the rectum as OAR with respect to acceptance for treatment planning?
- Q3. How confident are you in your final segmentation decisions (prostate CTV and rectum separately)?

The answer alternatives to Q1 and Q2 were:

- A. Excellent: Almost no modification necessary
- B. Good: Limited number of slices (2-5 slices) to be corrected
- C. Acceptable: Automatic segmentations can be used but require modification of several slices (> 5)
- D. Not acceptable: Useless, re-segmentation is needed

For A, B, C, D: Deviations < 1 mm in a slice were defined as negligible for prostate CTV. The corresponding value for the rectum was < 2mm.

The answer alternatives to Q3 were:

- a. Fully confident
- b. Somewhat confident
- c. Indeterminate/Indecisive
- d. Somewhat unconfident
- e. Totally unconfident

**2.3.2 Step 2 - Editing and assessment of deep learning segmentations with uncertainty map**

Edited segmentations for each patient case were exported and removed from Eclipse. The unedited DL-generated prostate CTV and rectum segmentations were again imported (same unedited segmentations as step 1). The uncertainty map for each segmentation was imported and displayed with the PET rainbow setting in Eclipse (similar to jet colormap, blue to red, range 0-0.5; see Fig.1). A minimum of four weeks from the last edited patient case in step 1 had to pass before the oncologist could begin step 2. Before the start of step 2, each oncologist answered the following question in free text:

"Do you think segmentation uncertainty information will provide you with useful information?"

The oncologist was asked to blend the DL segmentations and the colored uncertainty map while assessing and editing the segmentations in Eclipse, similar to performing clinical segmentation on PET-CT material. This was done using the exact instructions from step 1 and the same questions and answer scales.

After the oncologist had segmented all 35 patient cases in step 2, the following questions were answered in free text:

1. Do you think segmentation uncertainty information has provided you with useful information?
2. What were the benefits of segmentation uncertainty information?
3. What were the drawbacks of segmentation uncertainty information?
4. What do you think the potential of segmentation uncertainty information is?

Edited segmentations from step 2 were exported from Eclipse in DICOM format. Answers for each oncologist in steps 1 and 2 were recorded in personal and separate Google sheet documents. The oncologist did not have access to their answers from step 1 when performing step 2.

### 2.4 Data analysis

#### 2.4.1 Segmentation time

To detect differences, the recorded time needed for all patient cases was compared between step 1 and step 2 for all oncologists. For all statistical comparisons, a two-sided Wilcoxon signed rank test with a confidence level of 5% was used from the Python SciPy package (v.1.14.1) [27].

#### 2.4.2 Segmentation ratings

To determine if there was a difference between step 1 and step 2 in the oncologist rating for questions Q1, Q2, and Q3, a visual grading characteristics (VGC) analysis for ordinal data was conducted [28,29]. This analysis was performed using the software package "VGC Analyser 1.0 release 3" [30-32] with the following settings: bootstrap samples 2000, permutations 2000, 5% confidence level, trapezoid curve fitting, paired data, 35 cases for reference and test condition, respectively, fixed-reader analysis. VGC analysis is a non-parametric, rank-invariant method for evaluating visual grading data on image quality under two conditions (step 1 and step 2). It generates a VGC curve, from which the area under the curve (AUC) and its associated uncertainty can be determined. The AUC reflects the ability to differentiate between the two conditions.

#### 2.4.3 Segmentation editing and inter-observer analysis

All DICOM exported segmentations were converted to NIfTI format [33] using dcmrtstruct2nii [34] and read using Simple ITK [35] in Python. The following metrics were calculated for each segmentation and patient case to compare data from step 1, step 2, and the DL segmentations: Dice, normalized surface distance with 1 mm tolerance, further referred to as surface Dice [36], Hausdorff distance, volume difference, average surface distance, and total added path length with 1 mm tolerance.
A map of edited voxels was calculated between the two steps and DL segmentations by analyzing added or removed voxels. This map of edited voxels was used to calculate a histogram with the corresponding voxel uncertainty value in step 2. A visual analysis of the patient case with the largest difference between the two steps was also performed.

To investigate if there were any differences for individual oncologists between step 1 and step 2, the DL segmentations were used as reference segmentation, as they were the same in both steps. The metrics Dice, surface Dice, Hausdorff distance, volume difference, average surface distance, and total added path length were calculated for step 1 and step 2. The distributions from step 1 and step 2 for each metric were then compared for each oncologist using a two-sided Wilcoxon signed-rank test with a confidence level of 5%. Data from all oncologists was pooled for each step to assess inter-observer differences, and a Fligner-Killeen test [37] was performed to detect whether these distributions had equal variances, i.e., if the inter-observer differences were the same.

## 3. Results

### 3.1 Segmentation editing time

When including the uncertainty map, a decrease in editing time was observed among the oncologists. The median total organ segmentation time per patient case decreased in step 2 compared to step 1 with 0.7 minutes (-15%, p=0.013, median time 4.5 minutes for step 1), 1.0 minute (-14%, p=3.5*10^-5, median time 7.0 minutes for step 1), and 2 minutes (-33%, p=4.5*10^-6, median time 6.0 minutes for step 1) for the oncologists obsC, obsD, and obsE, respectively. Oncologist obsB reported an estimated 7-minute segmentation time per patient case in both step 1 and step 2.

### 3.2 Segmentation ratings

Results from the statistical analysis of the oncologist's visual grading rating showed differences between step 1 and step 2 for the prostate CTV and rectum (Table 1). The most common rating for each organ, for each oncologist, is reported for step 1 and step 2 in Table S1 in supplementary material, together with the lowest and highest ratings. All study answers from each oncologist in step 1 and 2 are also available as a supplementary file.

On a group level, a difference was detected regarding Q2, indicating that ratings of the unedited rectum segmentation were lower in step 2 compared to step 1 (AUC=0.44±0.02). However, no difference was detected for prostate CTV (Q1). Lower confidence in prostate CTV segmentation (Q3) was detected in step 2 (AUC=0.44±0.02) but none for the rectum (Q3). One out of four oncologists (obsE) had a statistically significant difference in the rating of prostate CTV (Q1), with a higher rating in step 2 (AUC=0.63±0.05). This oncologist also reported a slight increase in segmentation confidence for both the prostate CTV and rectum (Q3) in step 2 (AUC=0.56±0.03). Oncologist obsD had statistically significant lower ratings for Q2 in step 2 (rectum, AUC=0.37±0.04) and reported less confidence for both prostate CTV and rectum in step 2 (AUC=0.35±0.04 and AUC=0.41±0.04, respectively). Three oncologists reported in the free text that the uncertainty map provided helpful information and identified benefits. One reported drawback concerned inexperienced oncologists who may hesitate to edit outside the uncertainty map color region, potentially limiting necessary edits. One oncologist (obsD) did not find or report the uncertainty map helpful. Complete answers to free text questions before and after step 2 are available in the supplementary material.

### 3.3 Segmentation editing and inter-observer analysis

High-quality DL segmentations were obtained from nnUNet, and the oncologists needed limited editing. This was reflected in the high median Dice (0.97-0.99) and surface Dice scores (0.91-0.97) for the test data patient cases in step 1 against DL segmentations (Table 2).

The prostate CTV and rectum volume ratio comparing step 2 and step 1 among all patient cases in the test dataset were distributed around ratio 1; see supplementary Fig. S1 and S2 for data on each oncologist. High median Dice and surface Dice scores for the test data patient cases in step 1 against step 2 segmentations reflected differences between the steps where differences in the prostate CTV were larger than for the rectum (Table 2 and Fig. S3-S8 for Dice and Hausdorff in supplementary).

For three out of the four oncologists, the median Dice difference, calculated as median (step1vsDL-step2vsDL), for prostate CTV was negative, i.e., Dice was 0.01-0.02 higher in step 2, and less segmentation editing was performed, see Table S2 in the supplementary material. Less editing for the prostate CTV in step 2 was also verified in the decreased total added path length, where the median change among the observers ranged from 34 mm to 171 mm (p=0.00). There was also a statistically significant difference in all metrics for the oncologists as a group in step 2, showing fewer edits in step 2 with an improvement of 0.01 in Dice but no volume difference (p=0.89). For the rectum, no differences were detected in the metrics except for a small median volume change of 0.5 cm3; see Table S3 in the supplementary material. Greater adherence to the DL segmentation in step 2 for the prostate CTV was also seen in the increased Dice and surface Dice score in Table 2, comparing step 2 against DL to step 1 against DL.

For the prostate CTV inter-observer analysis, a difference in variance between step 1 and step 2 was detected for volume (3.64 cm3), average surface distance (0.02 mm), and total added path length (14649 mm), but not for Dice and surface Dice. No inter-observer differences were detected for the rectum. See supplementary Table S2 and Table S3 for the entire prostate CTV and rectum results, respectively.

Regarding the uncertainty map value, histogram data for all changed voxels over all patients in the test data for step 1 and step 2 also revealed differences between the steps. Histograms for each oncologist for the prostate CTV and rectum can be seen in Fig.2 and Fig.3, respectively. Note the smaller number of changed voxels in step 2 for all oncologists for prostate CTV (we remind the reader that the uncertainty map was not available to the oncologists in step 1). Specifically, 50% fewer voxel edits were recorded for the low uncertainty region 0.0-0.1 for prostate CTV, calculated over the whole patient case-cohort among the oncologists. There was no clear trend regarding the number of edited voxels between the steps for the rectum.

**3.4 Outlier analysis**

For prostate CTV, two patients-observer pairs (pat16 obsB, pat29 obsE) had a volume ratio between step 1 and step 2 larger than 1.1, and one patient-observer pair (pat31 obsE) had a volume ratio smaller than 0.91. For the rectum, one patient-observer pair (pat33 obsE) had a volume ratio above 1.25, and two patient-observer pairs (pat15 obsB, pat26 obsE) had a volume ratio less than 0.8. These outliers are also noticeable in the Dice and surface Dice plots comparing step 1 and step 2 for all oncologists for the prostate CTV and rectum (supplementary Fig.S5-S8).

The case analysis of the patient with the largest Dice difference between step 1 and step 2 for prostate CTV (pat16 obsB) revealed that fewer edits were made in step 2 with respect to the unedited DL segmentations. See Fig.4 and Fig.S9 supplementary material. The calculated Dice value between step 1 and DL segmentations and step 2 and DL segmentations was 0.91 and 0.98, respectively. A similar analysis of the patient case with the largest Dice difference for rectum (pat33 obsE) revealed that edited voxels were associated with larger uncertainty in step 2; see Fig.4 and Fig.S10 in the supplementary material.

## 4. Discussion

In this study, the impact of presenting DL segmentation uncertainty maps to four experienced radiation oncologists was qualitatively and quantitatively assessed with several metrics, subjective ratings, and free-text answers. The time to edit the DL segmentations was also assessed. Three of four oncologists had a decreased segmentation time of 1-2 minutes per patient case for segmentation of prostate CTV and rectum combined when the DL segmentations were presented with their uncertainty map compared to when the DL segmentations were presented alone.

The quantitative analysis of the DL segmentations and edited segmentations in step 1 without uncertainty maps showed that the average Dice range among the oncologists was 0.97-0.99 for both the prostate CTV and rectum. For surface Dice, the corresponding range was 0.91-0.97. This demonstrated that the segmentations produced by nnUNet were of high quality, and limited editing was needed for clinical use.

Subjective ratings of the DL segmentations were assessed in steps 1 and 2. On a group level (all oncologists), a small AUC decrease in the rectum rating was detected (Q2) and, similarly, in the decision confidence regarding prostate CTV (Q3). However, this deviation from an AUC of 0.5 (no detected difference) was slight and is not expected to make any substantial clinical difference. Individual oncologist variance existed; for example, the segmentation quality of the rectum varied between AUC 0.37 and 0.49. Oncologist obsD was the most influenced observer and reported lower segmentation confidence for the prostate CTV and rectum (AUC=0.35±0.04 and AUC=0.41±0.04, respectively). Interestingly, this observer did not report the uncertainty map to be helpful, but despite this, obsD was the most influenced observer in step 2.

The histogram analysis revealed that fewer voxels were edited in step 2 for prostate CTV for all oncologists and that low uncertainty regions contained the largest difference between step 1 and step 2 (Fig.2). We interpret this as the oncologists were influenced to trust the DL segmentation more in step 2 for regions with low uncertainty values. The results were not uniform among the oncologists for the rectum, and a trend was more challenging to identify (Fig.3). Less edited voxels in step 2 for prostate CTV were in line with the reported decrease in time spent editing the DL segmentations. These results also aligned with the higher median Dice value in step 2 than in step 1, calculated against the DL segmentations. Further, the decrease in total added path length in step 2 was consistent with the decrease in segmentation time, and these metrics have been shown to correlate [9].

The detected inter-observer differences between step 1 and step 2 for the prostate CTV were minor, where the variance in the average surface distance was clinically irrelevant. The decreased variance regarding volume and total added path length for the prostate CTV aligns with the other results demonstrating increased adherence to the DL segmentations in step 2.

Although the small difference in geometrical metrics, such as Dice between step 1 and step 2 over the whole patient case cohort in this study, might have limited clinical impact, the key findings of this study are the decreased segmentation time, and the tendency to avoid editing the DL segmentation in regions where the segmentation uncertainty is visibly low. Further, the DL uncertainty map can potentially aid in detecting patient outliers. These findings can significantly affect efficiency and trust in radiotherapy since DL-based segmentation is increasingly used. We believe this is further supported by the free-text answers, in which most oncologists noted that the provided information was helpful. Furthermore, the in-depth analysis of the patient case with the largest Dice difference between step 1 and step 2 for prostate CTV and rectum highlighted two interesting effects. The oncologist was more prone to follow the DL segmentation for the prostate case, which also defined the border towards larger uncertainty areas. This was in line with the reported results regarding less editing for the prostate CTV in step 2. For the rectum case, the volume with the largest uncertainty likely influenced the oncologist, and the additional volume was included. In this case, more editing was done in step 2.

Published results on similar studies are limited, as we are the first to assess the impact of segmentation uncertainty in a clinical setting. Prior work on mapping segmentation uncertainty in the lungs, head and neck, and prostate has highlighted the need to assess its clinical impact (15, 17, 18, 21). In De Biase and Ziegfeld's [21] work, a deep learning method was created and presented through a graphical user interface to generate tumor probability maps for PET-CT imaging, replacing fixed contours and allowing users to interact with the maps. The study also included an assessment of the user experience. The results indicated that oncologists preferred the rainbow colormap, similar to the colormap adopted in this study. In line with the results of our study, they found that the probability map was intuitive, timesaving, clinically feasible, and was preferred over single contours.

Another recent study assessing the radiation oncologists' and radiation therapists' preferences for uncertainty visualization methods concluded that binary voxel-level uncertainty visualization was preferred [18] above the continuous visualization scale used in our study. This result could be an essential factor for future studies in this area.

To assess segmentation uncertainty in our study, we computed the voxel-wise standard deviation (std) across a 10-fold cross-validation. This captures a mixture of two types of uncertainty. Epistemic uncertainty reflects the model's inability to fully capture all possible patterns due to insufficient data or model limitations, i.e. systematic errors, such as model choice and fitting and variation from the cross-validation training data folds. On the other hand, aleatoric uncertainty arises from inherent characteristics of the dataset, such as noise, ambiguous boundaries, or inconsistencies in ground truth annotations [38]. The method presented in this work does not distinguish between the two uncertainty types making this approach a practical, blended measure of predictive uncertainty.

Before the study, we hypothesized that the high uncertainty areas would be the parts the oncologist focused on in step 2. However, the results for prostate CTV indicate that it instead was the low uncertainty regions that were most influential, which was not expected. The corresponding results for the rectum were not as consistent as for the prostate CTV.

The study setup relied on contouring 35 patient cases at two-time points with four weeks in between. The detected changes between step 1 and step 2 can have been influenced by intra-observer variation, which was not assessed in this study. Therefore, the effects cannot be isolated here. The high quality of the produced DL segmentations before editing could also have contributed to automation bias, meaning that the oncologists could have over-relied on the DL segmentation. However, the prostate CTV segmentation analysis indicated that all oncologists segmented fewer voxels in step 2. Another limitation might be that the order of step 1 and step 2 was not randomized, and step 2 always followed step 1. Any bias originating from the repetitive task of segmenting the patient case a second time, always with the uncertainty map present, cannot be excluded. Further, the robustness of the results is limited by the small number of oncologists and the results should be validated in further studies.

With regards to the study results, we believe that visualization of the DL uncertainty can play an important role when DL segmentations need to be assessed more efficiently, such as in adaptive radiotherapy workflows or an MR-linac setting. This is especially important when DL segmentation becomes a clinical standard [14,16].

## 5. Conclusion

In conclusion, when DL uncertainty information is presented to experienced radiation oncologists, it can influence their decision-making, quality perception, and confidence in the DL segmentations. When the uncertainty map was presented, regions with low uncertainty were less likely to be edited, indicating increased reliance on the model's predictions. Additionally, uncertainty maps can improve efficiency by reducing segmentation time. DL-based segmentation uncertainty can be a valuable tool in clinical practice, enhancing the efficiency of radiotherapy planning.

**Tables**

**Table 1.** AUC is calculated for each question where data between the two conditions, step 1 and step 2, have been used. AUC reflects the difference between the two conditions where AUC < 0.5 refers to higher ratings for step 1 than step 2, and vice versa. Values provided are AUC±1 standard deviation [Asymmetric 95% confidence interval] and (p-value). * and bold font indicates statistically significant results (p<0.05). Statistics were calculated for trapezoid curve fitting.

| Oncologist | Q1 | Q2 | Q3 Prostate CTV | Q3 Rectum |
|---|---|---|---|---|
| obsB | 0.43±0.04 [0.34, 0.51] (0.18) | 0.49±0.04 [0.42, 0.58] (0.91) | 0.45±0.05 [0.35, 0.55] (0.30) | 0.47±0.05 [0.39, 0.56] (0.54) |
| obsC | 0.52±0.06 [0.40, 0.64] (0.72) | 0.44±0.04 [0.36, 0.52] (0.13) | 0.41±0.05 [0.30, 0.51] (0.10) | 0.45±0.03 [0.38, 0.51] (0.21) |
| obsD | 0.53±0.03 [0.47, 0.60] (0.22) | **0.37±0.04 [0.28, 0.45] (0.01) *** | **0.35±0.04 [0.27, 0.43] (0.00) *** | **0.41±0.04 [0.34, 0.48] (0.01) *** |
| obsE | **0.63±0.05 [0.54, 0.73] (0.01) *** | 0.46±0.04 [0.39, 0.54] (0.45) | **0.56±0.03 [0.50, 0.63] (0.02) *** | **0.56±0.03 [0.50, 0.63] (0.03) *** |
| All oncologists as a group | 0.53±0.02 [0.48, 0.57] (0.23) | **0.44±0.02 [0.40, 0.48] (0.01) *** | **0.44±0.02 [0.40, 0.49] (0.01) *** | 0.47±0.02 [0.44, 0.50] (0.09) |

**Table 2.** Median Dice and median surface Dice±1 standard deviation [min, max] calculated for test data patient cases after editing in step 1 against DL segmentations (upper part), step 2 against DL segmentations (middle part), and after editing in step 2 against step 1 segmentations (lower part). Data shows values for the prostate CTV and the rectum for oncologist obsB-obsE.

| **Step 1 against DL segmentations** | **Prostate CTV** | | **Rectum** | |
|---|---|---|---|---|
| Oncologist | **Dice** | **Surface Dice** | **Dice** | **Surface Dice** |
| obsB | 0.98±0.02 [0.90, 0.99] | 0.95±0.07 [0.67, 0.99] | 0.99±0.02 [0.89, 1.00] | 0.96±0.04 [0.83, 0.99] |
| obsC | 0.98±0.02 [0.91, 1.00] | 0.93±0.05 [0.80, 1.00] | 0.98±0.02 [0.92, 1.00] | 0.94±0.04 [0.84, 0.99] |
| obsD | 0.98±0.02 [0.89, 1.00] | 0.93±0.06 [0.68, 1.00] | 0.98±0.02 [0.88, 1.00] | 0.94±0.04 [0.83, 1.00] |
| obsE | 0.97±0.02 [0.90, 1.00] | 0.91±0.07 [0.71, 1.00] | 0.99±0.03 [0.88, 1.00] | 0.97±0.04 [0.83, 1.00] |
| **Step 2 against DL segmentations** | | | | |
| Oncologist | **Dice** | **Surface Dice** | **Dice** | **Surface Dice** |
| obsB | 0.98±0.01 [0.92, 0.99] | 0.96±0.03 [0.81, 0.99] | 0.99±0.01 [0.96, 1.00] | 0.97±0.03 [0.85, 1.00] |
| obsC | 0.99±0.02 [0.93, 1.00] | 0.96±0.05 [0.82, 1.00] | 0.98±0.02 [0.92, 1.00] | 0.94±0.04 [0.81, 1.00] |
| obsD | 0.99±0.01 [0.92, 0.99] | 0.96±0.03 [0.81, 0.98] | 0.98±0.03 [0.89, 1.00] | 0.94±0.04 [0.81, 0.99] |
| obsE | 1.00±0.01 [0.93, 1.00] | 0.99±0.03 [0.83, 1.00] | 0.99±0.02 [0.90, 1.00] | 0.98±0.04 [0.80, 1.00] |
| **Step 1 against step 2 segmentations** | | | | |
| Oncologist | **Dice** | **Surface Dice** | **Dice** | **Surface Dice** |
| obsB | 0.98±0.01 [0.92, 0.99] | 0.96±0.05 [0.70, 0.99] | 0.99±0.02 [0.88, 1.00] | 0.97±0.03 [0.84, 1.00] |
| obsC | 0.98±0.01 [0.94, 1.00] | 0.93±0.04 [0.86, 1.00] | 0.99±0.01 [0.94, 1.00] | 0.96±0.03 [0.88, 1.00] |
| obsD | 0.98±0.01 [0.96, 0.99] | 0.94±0.03 [0.85, 0.99] | 0.99±0.02 [0.91, 1.00] | 0.96±0.03 [0.85, 1.00] |
| obsE | 0.97±0.02 [0.93, 1.00] | 0.90±0.06 [0.78, 1.00] | 0.99±0.03 [0.87, 1.00] | 0.98±0.04 [0.83, 1.00] |

**Figures**

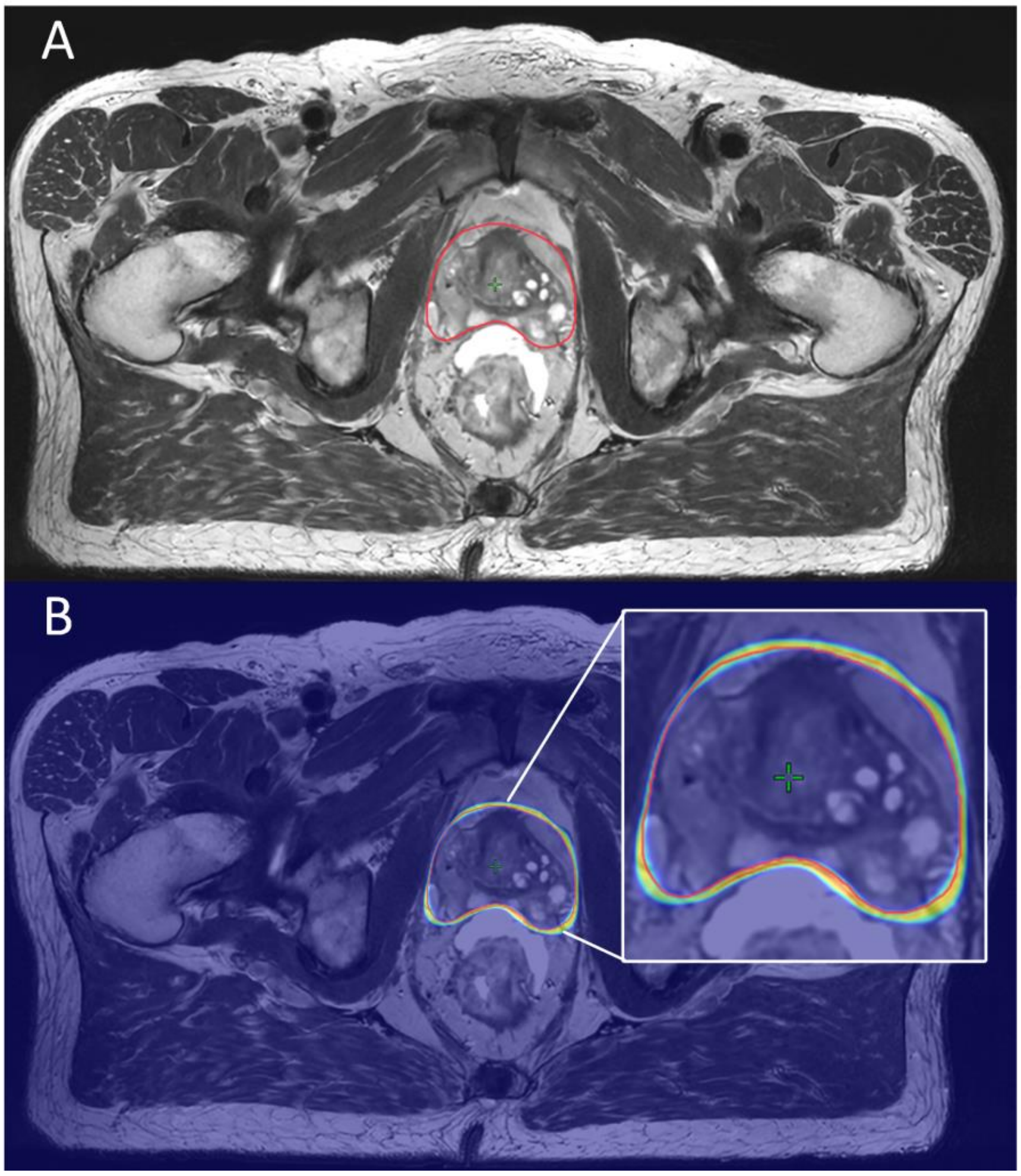

**Fig.1**. A. Example of prostate CTV data for one patient case (transverse view) in step 1 and DL segmentation in red. B. The corresponding uncertainty map used in step 2 with a 50 % blend with the transverse view in A.

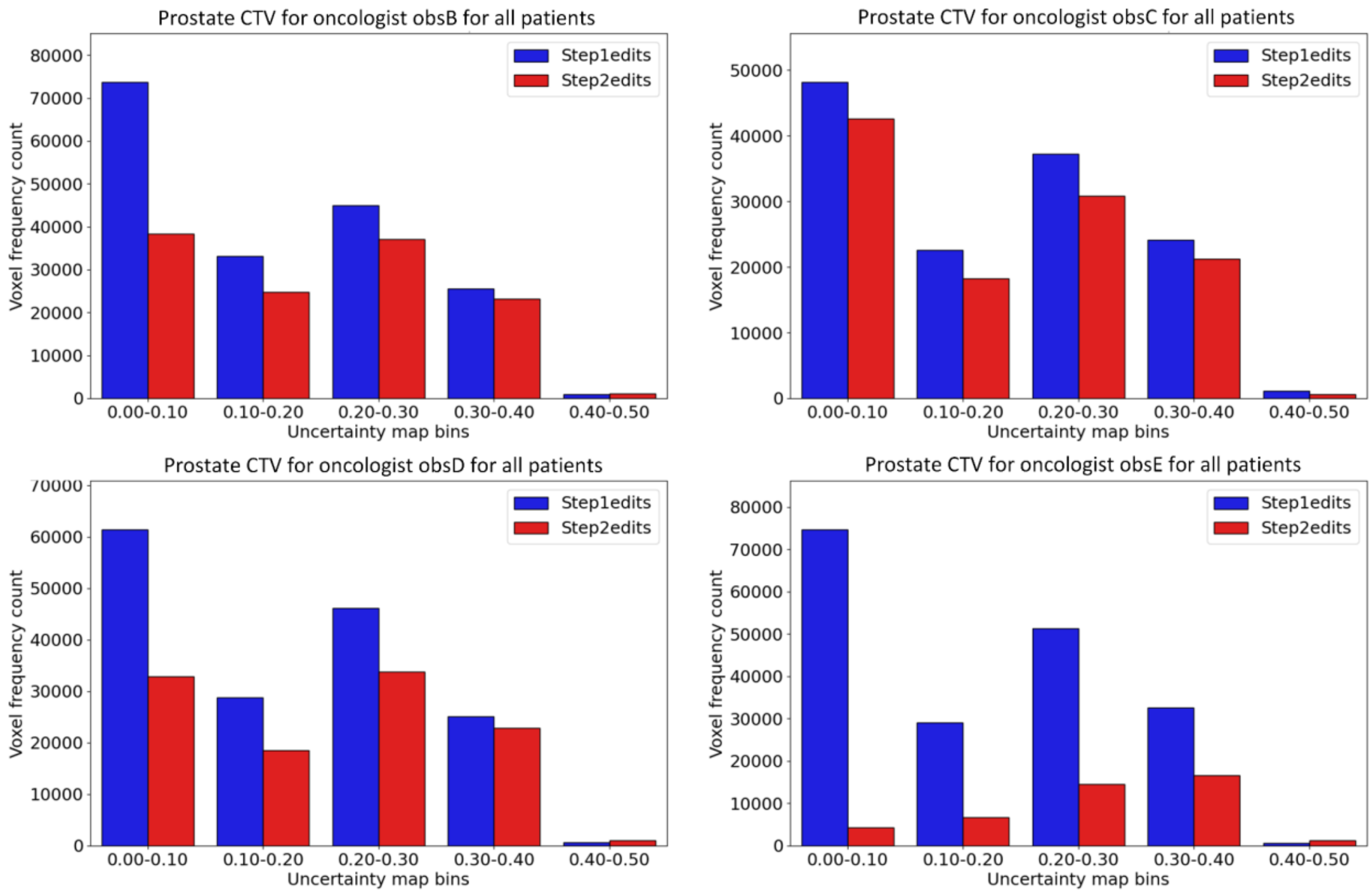

**Fig.2**. Histogram over changed voxels for prostate CTV in all patients in step 1 and step 2 for all oncologists individually. Note fewer edits in step 2 for all oncologists, especially in the low-value region (0.00-0.10).

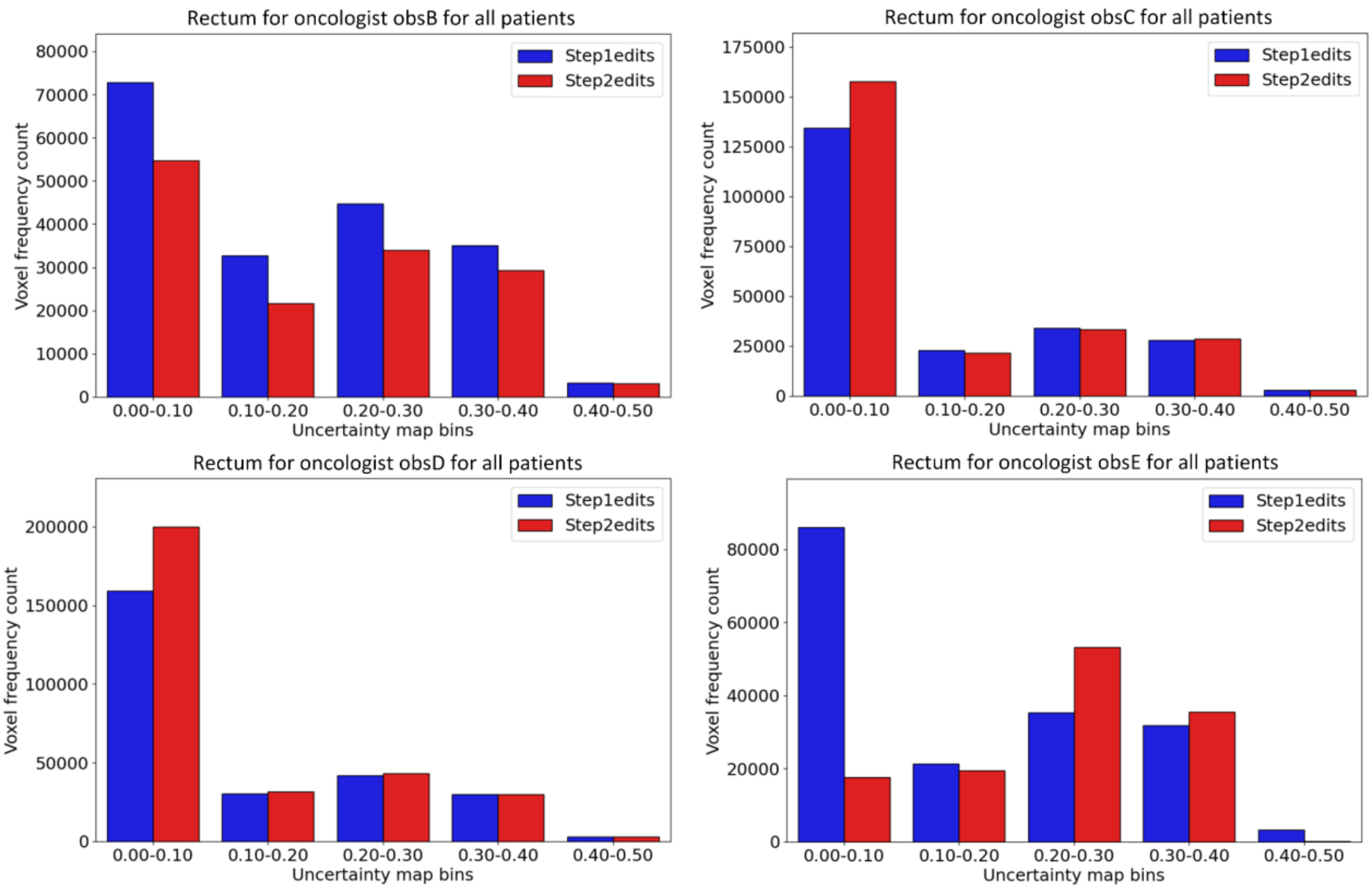

**Fig.3**. Histogram over changed voxels for rectum in all patients in step 1 and step 2 for all oncologists individually. There was no clear trend regarding the number of edited voxels between the steps for the rectum.

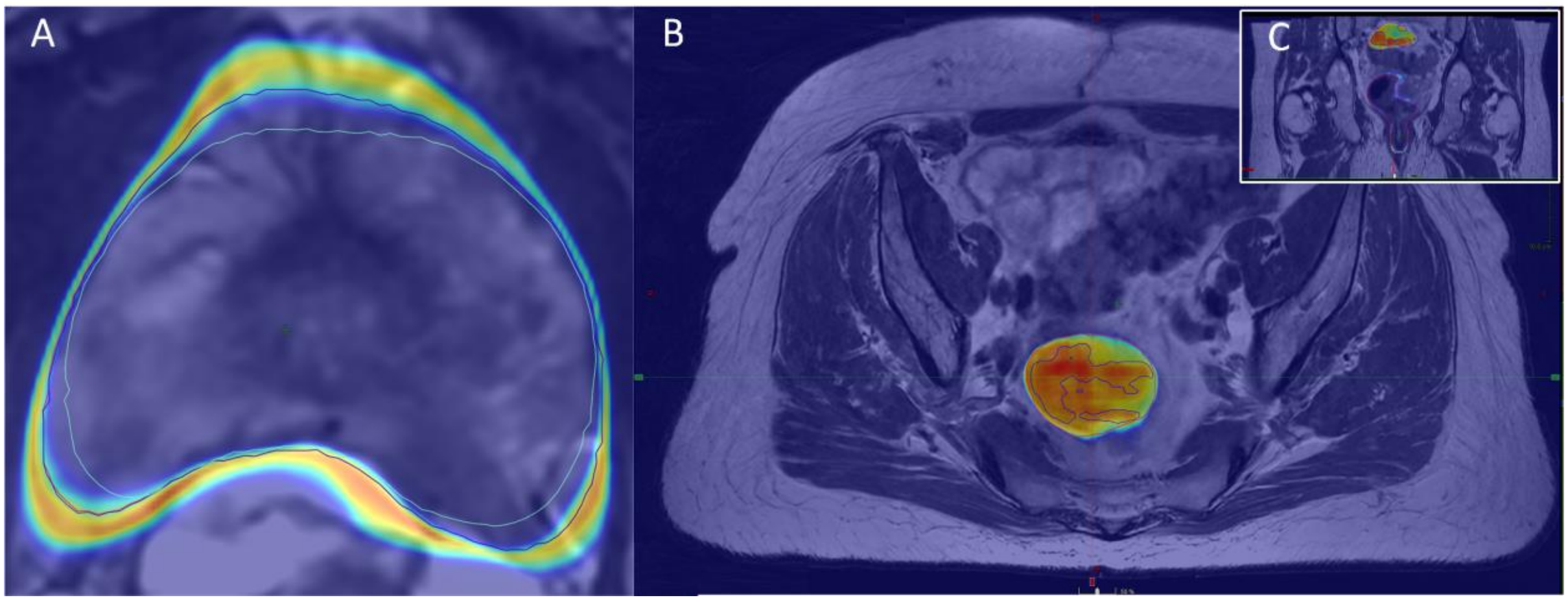

**Fig.4**. A. Prostate CTV segmentation of patient-observer pair 16 obsB. Blue segmentation is DL segmentation, green segmentation is from step 1, and red is from step 2 (similar to DL segmentation). B. Rectum segmentation of patient-observer pair 33 obsE. The same color codes for segmentation are used in the prostate case (A). A sub-volume with large uncertainty is displayed, and this volume was included in step 2 but not in step 1, explaining the volume increase. Partial segmentation of this volume can also be seen in the DL segmentation. C. Coronal view of the patient displayed in B.

## Supplementary material

### Study setup

Internal documentation and methods used in Eclipse during the study are available in the file "Supplementary_Data import and export instructions v2_public.pdf".

### Oncologist organ rating study raw data

All oncologist raw data ratings for each question and each subject can be found in the file "Supplementary_ObserverRawData.zip". This zip file contains a copy of the Google Docs document each oncologist used in the study for step 1 and step 2.

### Oncologist organ ratings

**Table S1**. The most common rating was reported for each question/organ, for each oncologist in step 1 and step 2, followed by the highest and lowest in brackets. The number in parentheses following the rating letter is the frequency.

| Oncologist | Q1 | | Q2 | | Q3 CTV | | Q3 Rectum | |
|---|---|---|---|---|---|---|---|---|
| | Step 1 | Step 2 | Step 1 | Step 2 | Step 1 | Step 2 | Step 1 | Step 2 |
| obsB | B (22) [B-C] | C (18) [B-C] | B (17) [A-C] | B (22) [A-C] | a (21) [a-c] | b (18) [a-b] | a (23) [a-b] | a (21) [a-c] |
| obsC | B (24) [A-C] | B (15) [A-C] | B (22) [A-C] | B (19) [A-C] | a (19) [a-c] | b (21) [a-b] | a (33) [a-c] | a (29) [a-c] |
| obsD | B (30) [A-D] | B (32) [A-C] | B (28) [B-C] | B (19) [B-D] | b (23) [a-c] | b (29) [a-c] | a (25) [a-b] | a (19) [a-c] |
| obsE | A (17) [A-C] | A (26) [A-C] | A (27) [A-C] | A (24) [A-C] | a (29) [a-b] | a (33) [a-b] | a (30) [a-b] | a (34) [a-b] |

**Oncologist inter-observer differences**

**Table S2**. The median of the differences, i.e., median(step 1-step 2), ± 1 std [minimum, maximum] (p-value) for multiple metrics for the prostate structure for all observers individually. Step 1 and step 2 data were calculated with respect to the reference nnUNet structure. Wilcoxon Signed-Rank Test was used to determine statistical significance for each observer. For "All oncologists as a group", all data from step 1 and step 2 have been pooled for all observers. A Fligner-Killeen test for homogeneity of variances across all the observers was used to determine if the inter-observer differences between step 1 and step 2 was statistically significant. The difference in variance var(step 1)-var(step 2) is reported as varDiff after the p-value in "Inter-observer difference". ***** and bold font represent statistically significant results.

| | DSC | Surface DSC | HD (mm) | HD95 (mm) | Volume difference (cm3) | Average surface distance ref2obs (mm) | Total Added Path length (mm) |
|---|---|---|---|---|---|---|---|
| obsB | **-0.01±0.01 [-0.07, 0.02] (p=0.00) *** | **-0.02±0.05 [-0.24, 0.06] (p=0.00) *** | **0.17±0.86 [-0.57, 2.86] (p=0.00) *** | **0.27±0.55 [-0.94, 1.45] (p=0.00) *** | **0.49±1.12 [-1.59, 5.29] (p=0.00) *** | **0.05±0.15 [-0.17, 0.78] (p=0.00) *** | **33.75±124.32 [-88.13, 623.50] (p=0.00) *** |
| obsC | -0.00±0.02 [-0.04, 0.03] (p=0.31) | -0.00±0.05 [-0.09, 0.10] (p=0.49) | 0.00±1.97 [-3.53, 7.46] (p=0.91) | 0.00±0.98 [-1.80, 2.10] (p=0.26) | 0.00±1.11 [-3.30, 2.28] (p=0.79) | 0.00±0.11 [-0.27, 0.23] (p=0.49) | 0.00±99.15 [-186.11, 206.27] (p=0.67) |
| obsD | **-0.01±0.01 [-0.03, 0.03] (p=0.00) *** | **-0.02±0.05 [-0.13, 0.10] (p=0.00) *** | 0.34±1.56 [-5.00, 4.16] (p=0.08) | **0.40±0.83 [-2.10, 1.88] (p=0.01) *** | **-0.61±1.04 [-4.52, 1.14] (p=0.00) *** | **0.04±0.10 [-0.24, 0.32] (p=0.00) *** | **23.44±98.35 [-254.56, 289.25] (p=0.00) *** |
| obsE | **-0.02±0.02 [-0.07, 0.01] (p=0.00) *** | **-0.07±0.06 [-0.22, 0.01] (p=0.00) *** | **1.20±1.84 [-2.50, 5.00] (p=0.00) *** | **1.48±1.00 [-0.47, 3.13] (p=0.00) *** | 0.04±2.20 [-5.17, 4.61] (p=0.88) | **0.17±0.17 [-0.04, 0.53] (p=0.00) *** | **171.11±140.79 [-21.10, 596.31] (p=0.00) *** |
| All oncologists as a group | **-0.01±0.02 [-0.07, 0.03] (p=0.00) *** | **-0.03±0.06 [-0.24, 0.10] (p=0.00) *** | **0.35±1.73 [-5.00, 7.46] (p=0.00) *** | **0.40±0.99 [-2.10, 3.13] (p=0.00) *** | 0.01±1.54 [-5.17, 5.29] (p=0.89) | **0.06±0.15 [-0.27, 0.78] (p=0.00) *** | **36.10±131.11 [-254.56, 623.50] (p=0.00) *** |
| Inter-observer difference | **p=0.00*, varDiff=0.00** | **p=0.00*, varDiff=0.00** | p=0.08, varDiff=0.05 | p=0.21, varDiff=-0.06 | **p=0.00*, varDiff=3.64** | **p=0.00*, varDiff=0.02** | **p=0.00*, varDiff=14649.08** |

**Table S3**. Median of the differences, i.e. median(step 1-step 2), ± 1 std [minimum, maximum] (p-value) for multiple metrics for the rectum structure for all observers individually. Step 1 and step 2 data were calculated against the reference nnUNet structure. Wilcoxon Signed-Rank test was used to determine statistical significance for each observer. For "All oncologists as a group", all data from step 1 and step 2 have been pooled for all observers. A Fligner-Killeen test for homogeneity of variances across all the observers was used to determine if the inter-observer differences between step 1 and step 2 were statistically significant. The difference in variance var(step 1)-var(step 2) is reported as varDiff after the p-value in "Inter-observer difference". ***** and bold font represent statistically significant results.

| | DSC | SurfaceDSC | HD (mm) | HD95 (mm) | Volume difference (cm3) | Average surface distance ref2obs (mm) | Total Added Path length (mm) |
|---|---|---|---|---|---|---|---|
| obsB | -0.00+/-0.02 [-0.10, 0.02] (p=0.05) | -0.01+/-0.03 [-0.14, 0.05] (p=0.09) | 0.04+/-4.16 [-4.48, 17.95] (p=0.37) | 0.00+/-3.22 [-3.95, 17.45] (p=0.24) | **-0.91+/-2.49 [-13.11, 1.58] (p=0.00) *** | 0.00+/-0.15 [-0.61, 0.20] (p=0.52) | 0.00+/-68.78 [-218.46, 131.26] (p=0.27) |
| obsC | 0.00+/-0.02 [-0.05, 0.05] (p=0.10) | 0.00+/-0.03 [-0.07, 0.06] (p=0.21) | 0.00+/-4.87 [-21.79, 8.53] (p=0.23) | **0.00+/-4.21 [-19.71, 7.50] (p=0.03) *** | **-0.61+/-2.39 [-6.75, 6.55] (p=0.04) *** | **-0.08+/-0.65 [-3.16, 0.98] (p=0.01) *** | -22.50+/-141.32 [-438.80, 414.89] (p=0.07) |
| obsD | 0.00+/-0.02 [-0.02, 0.09] (p=0.54) | 0.00+/-0.03 [-0.03, 0.15] (p=0.38) | 0.00+/-5.78 [-22.72, 9.39] (p=0.44) | 0.00+/-4.45 [-15.00, 9.00] (p=0.14) | -0.20+/-2.69 [-12.94, 3.03] (p=0.24) | -0.01+/-0.58 [-2.03, 1.45] (p=0.10) | -7.50+/-156.86 [-766.02, 146.27] (p=0.20) |
| obsE | -0.00+/-0.03 [-0.12, 0.06] (p=0.05) | -0.00+/-0.04 [-0.17, 0.06] (p=0.06) | 0.19+/-10.07 [-7.24, 43.43] (p=0.12) | **0.00+/-7.53 [-2.50, 32.92] (p=0.01) *** | -0.02+/-7.81 [-22.13, 38.89] (p=0.19) | **0.01+/-0.85 [-0.03, 5.15] (p=0.00) *** | **6.56+/-43.82 [-20.16, 224.56] (p=0.00) *** |
| All oncologists as a group | 0.00+/-0.02 [-0.12, 0.09] (p=0.48) | 0.00+/-0.03 [-0.17, 0.15] (p=0.46) | 0.00+/-6.90 [-22.72, 43.43] (p=0.84) | 0.00+/-5.39 [-19.71, 32.92] (p=0.73) | **-0.51+/-4.50 [-22.13, 38.89] (p=0.00) *** | 0.00+/-0.63 [-3.16, 5.15] (p=0.09) | 0.00+/-115.75 [-766.02, 414.89] (p=0.21) |
| Inter-observer difference | p=0.79, varDiff=0.00 | p=0.65, varDiff=-0.00 | p=0.98, varDiff=-2.41 | p=0.18, varDiff=2.37 | p=0.76, varDiff=-2.92 | p=0.17, varDiff=-0.19 | p=0.22, varDiff=-12575.43 |

**Inter-observer differences**

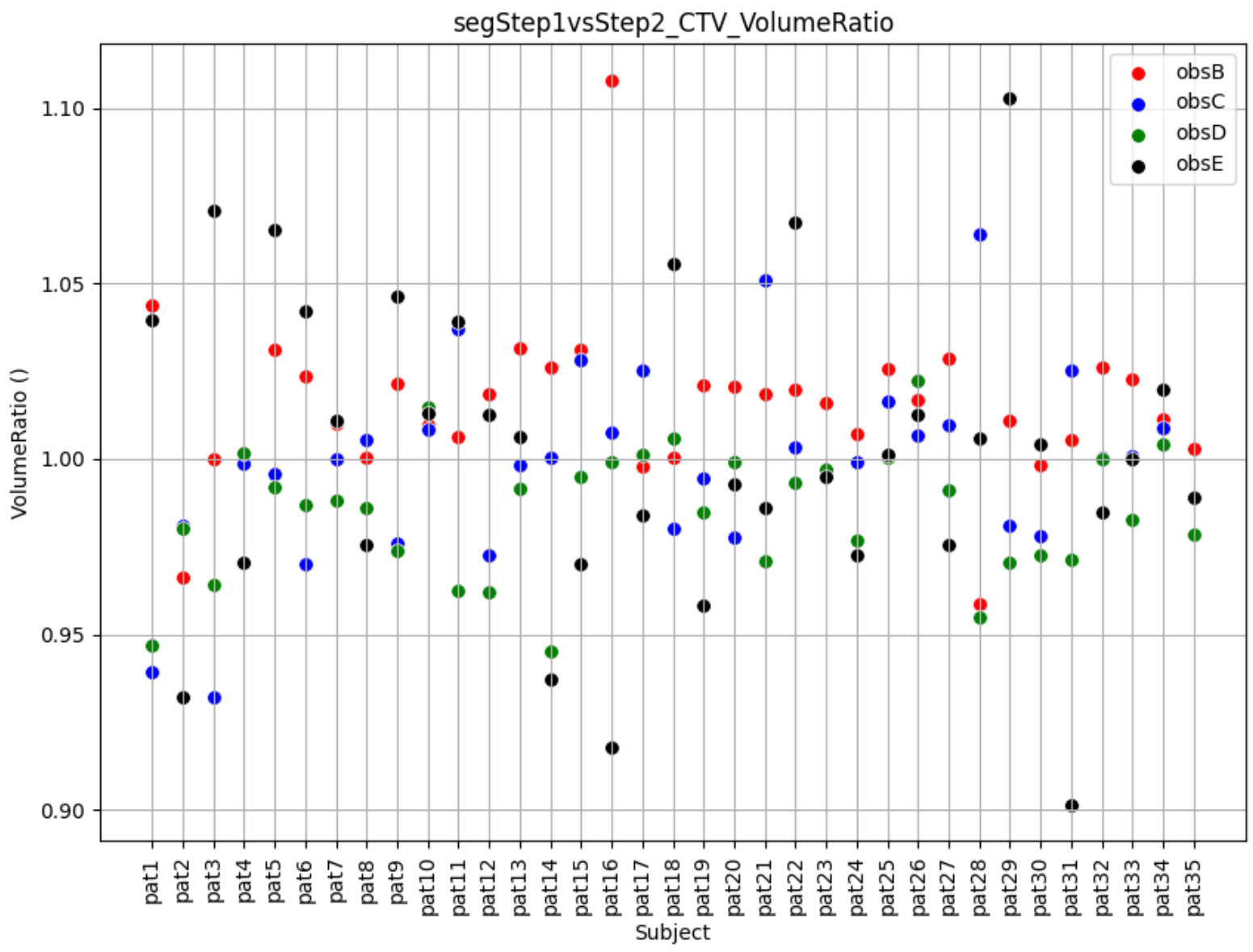

**Fig.S1**. Volume ratio step 2/step 1 among all patients for all observers in the test dataset for CTV.

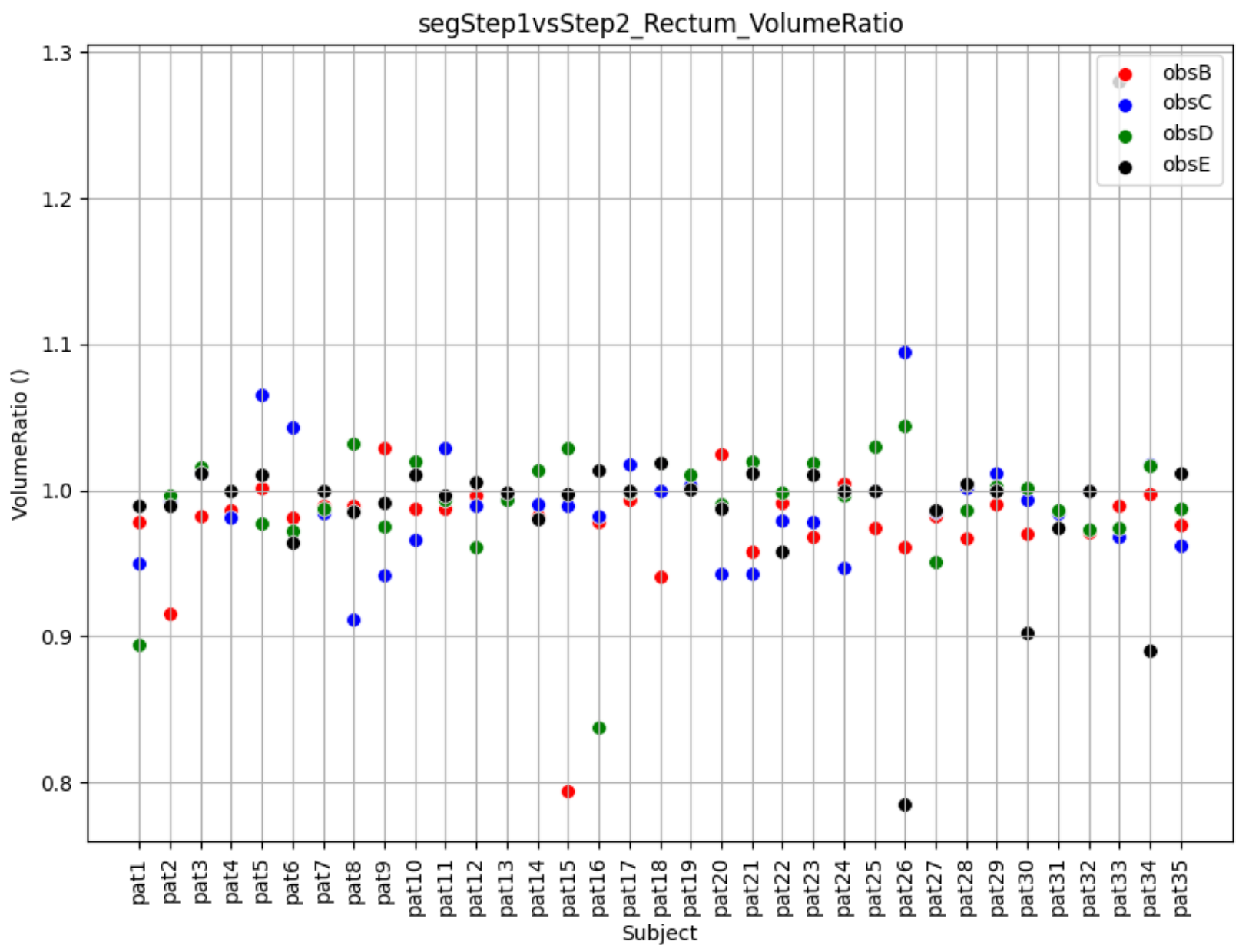

**Fig.S2**. Volume ratio step 2/step 1 among all patients for all observers in the test dataset for the rectum.

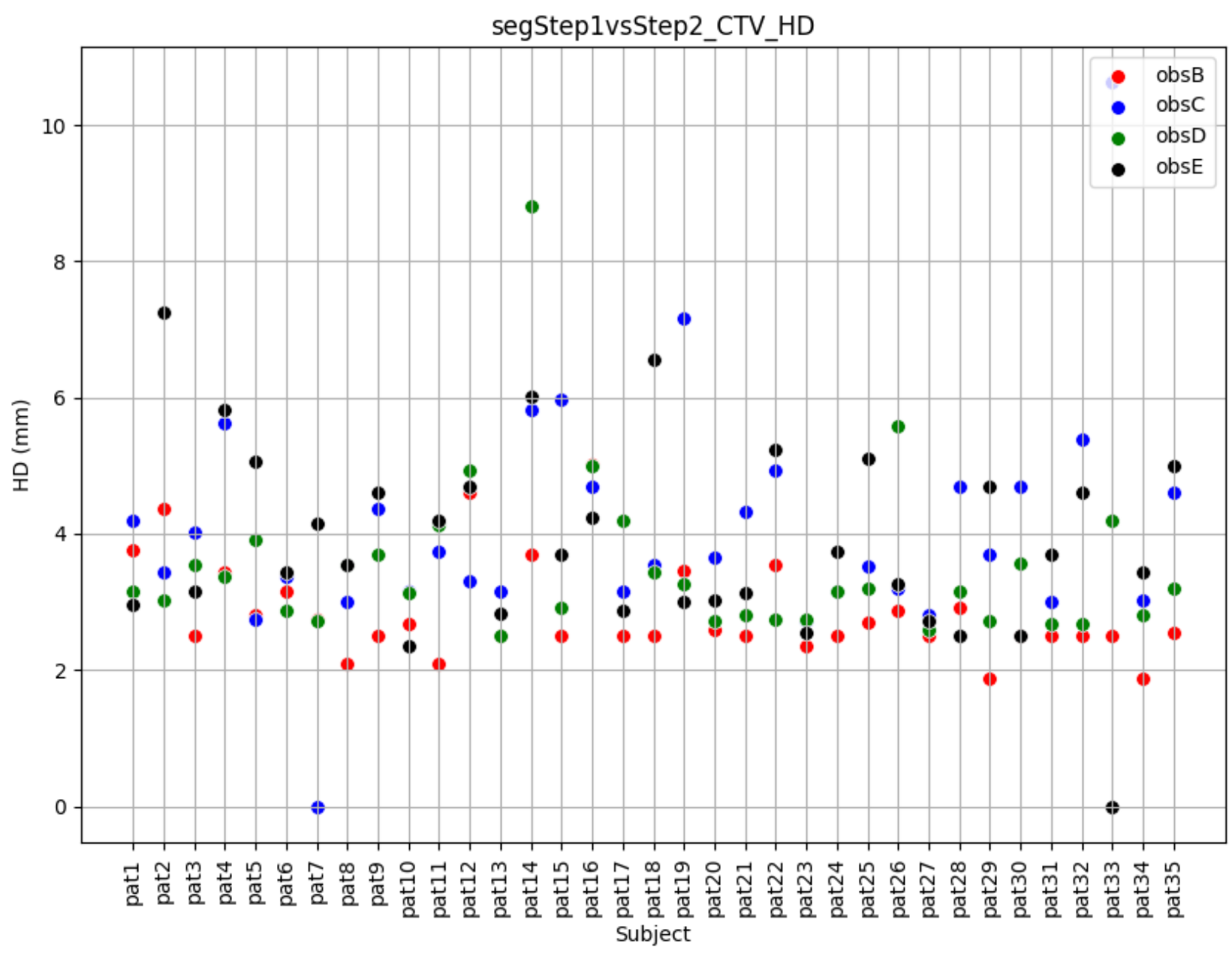

**Fig.S3**. Hausdorff distance between step 2 and step 1 among all patients for all observers in the test dataset for the prostate.

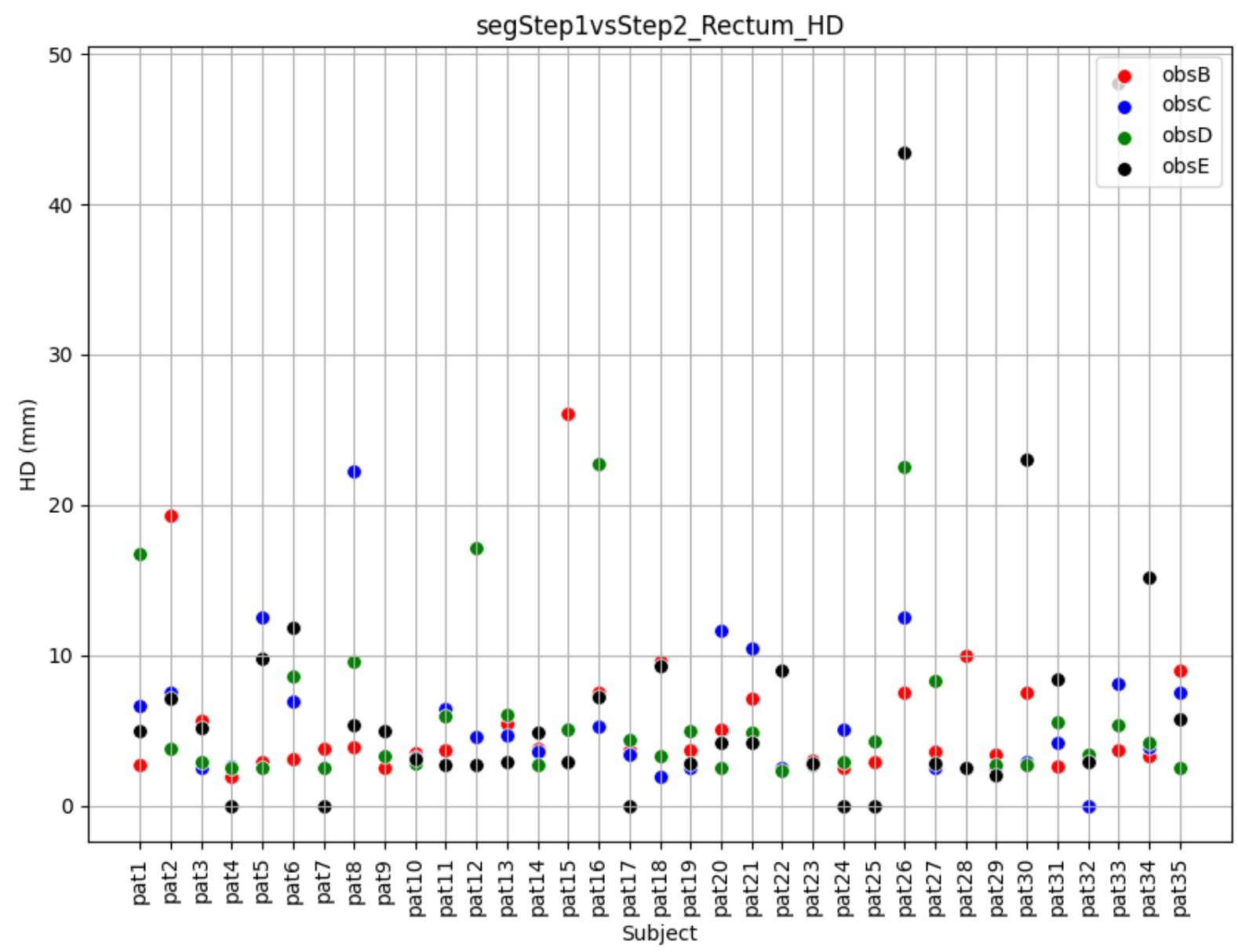

**Fig.S4**. Hausdorff distance between step 2 and step 1 among all patients for all observers in the test dataset for the rectum.

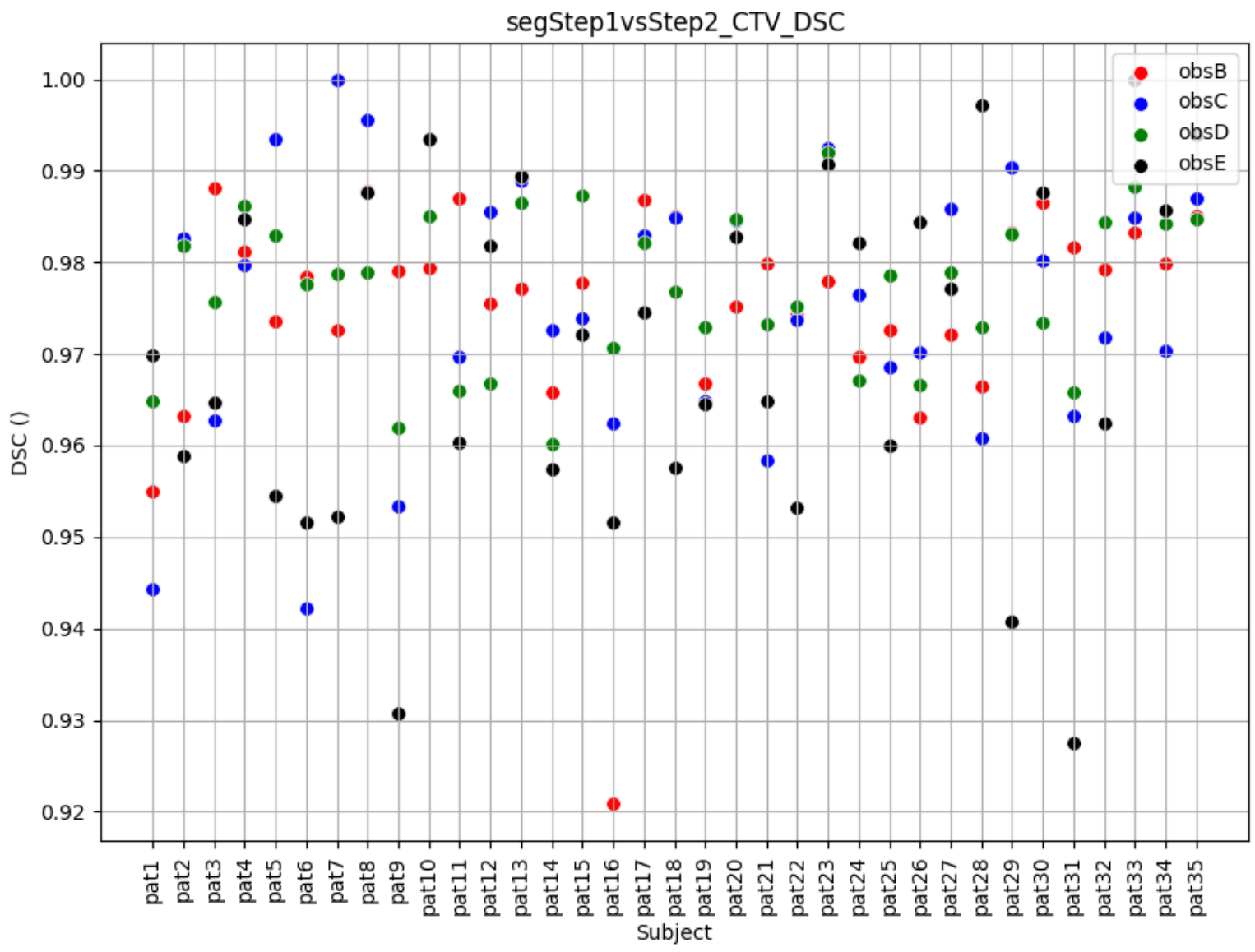

**Fig.S5**. Dice and between step 1 and step 2 for the prostate.

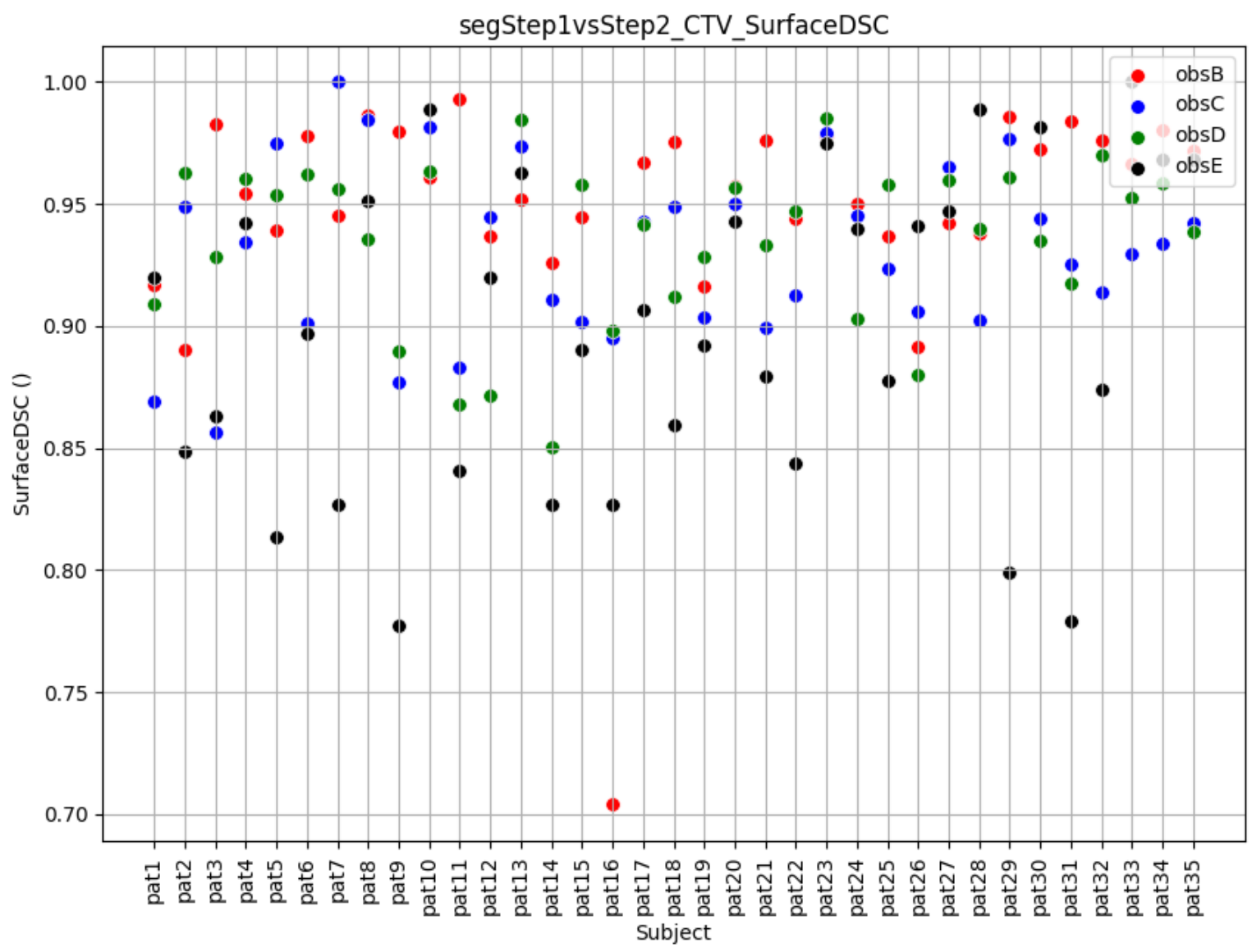

**Fig.S6**. Surface Dice between step 1 and step 2 for the prostate.

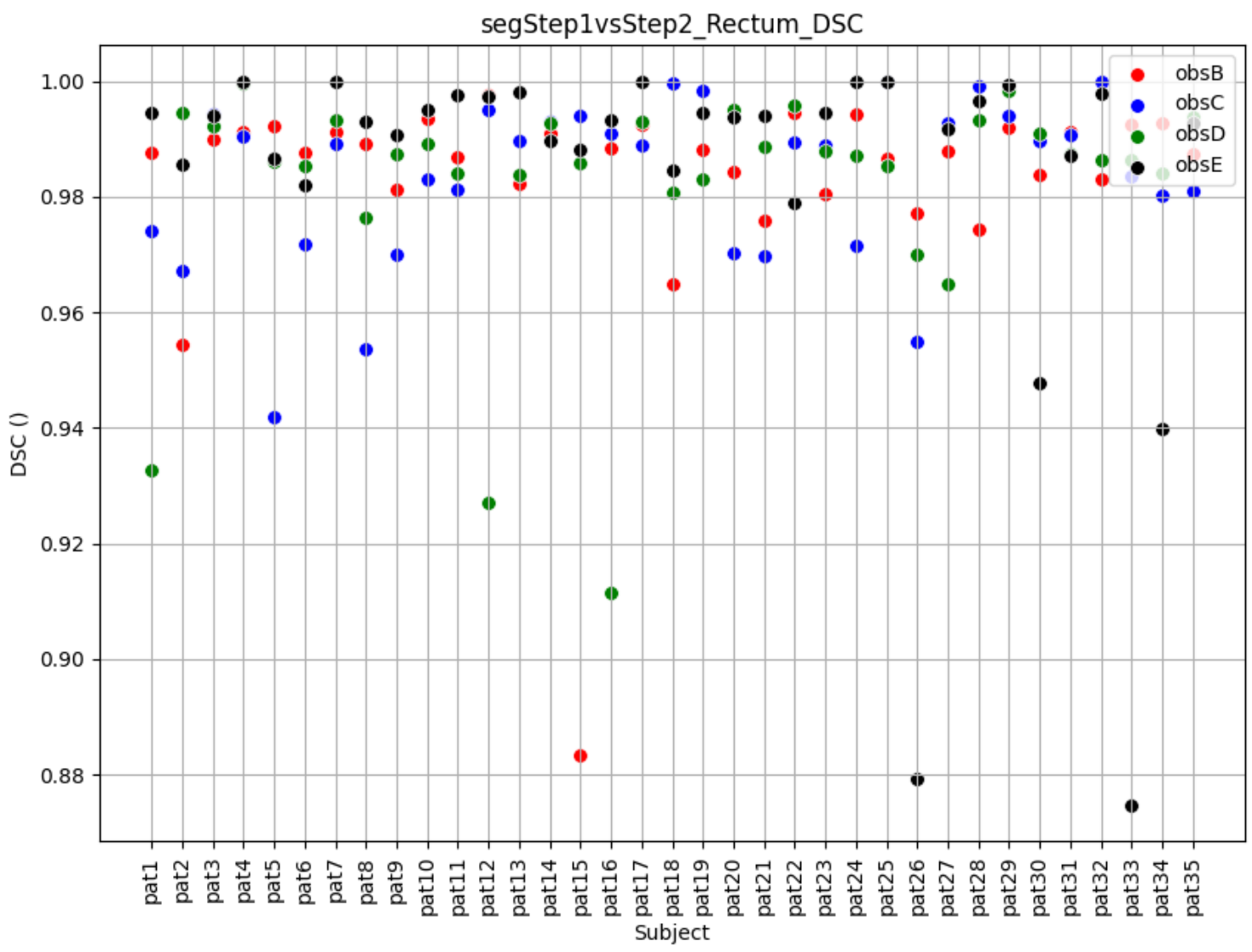

**Fig.S7**. Dice between step 1 and step 2 for the rectum.

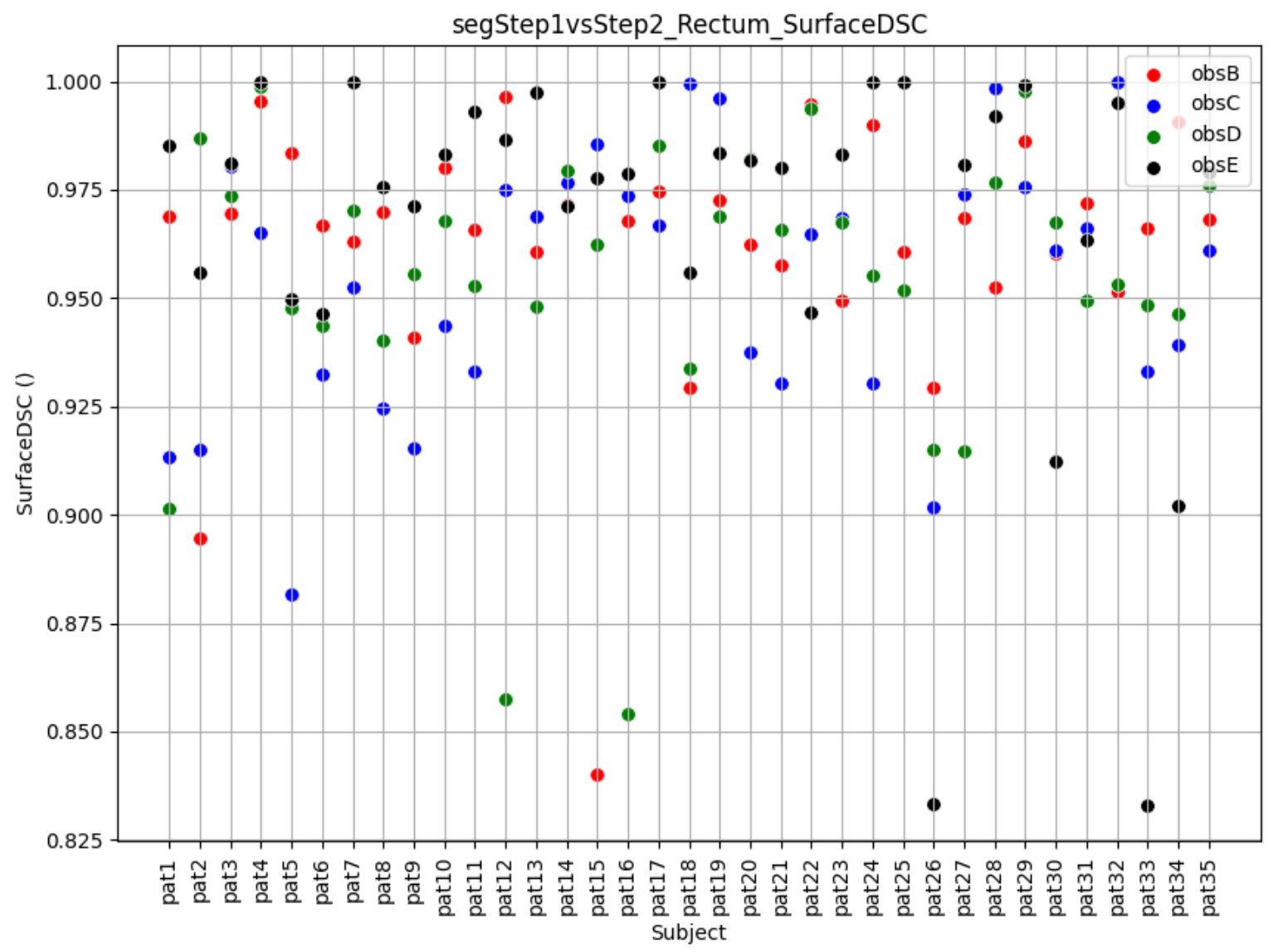

**Fig.S8**. Surface Dice between step 1 and step 2 for the rectum.

**Outlier analysis**

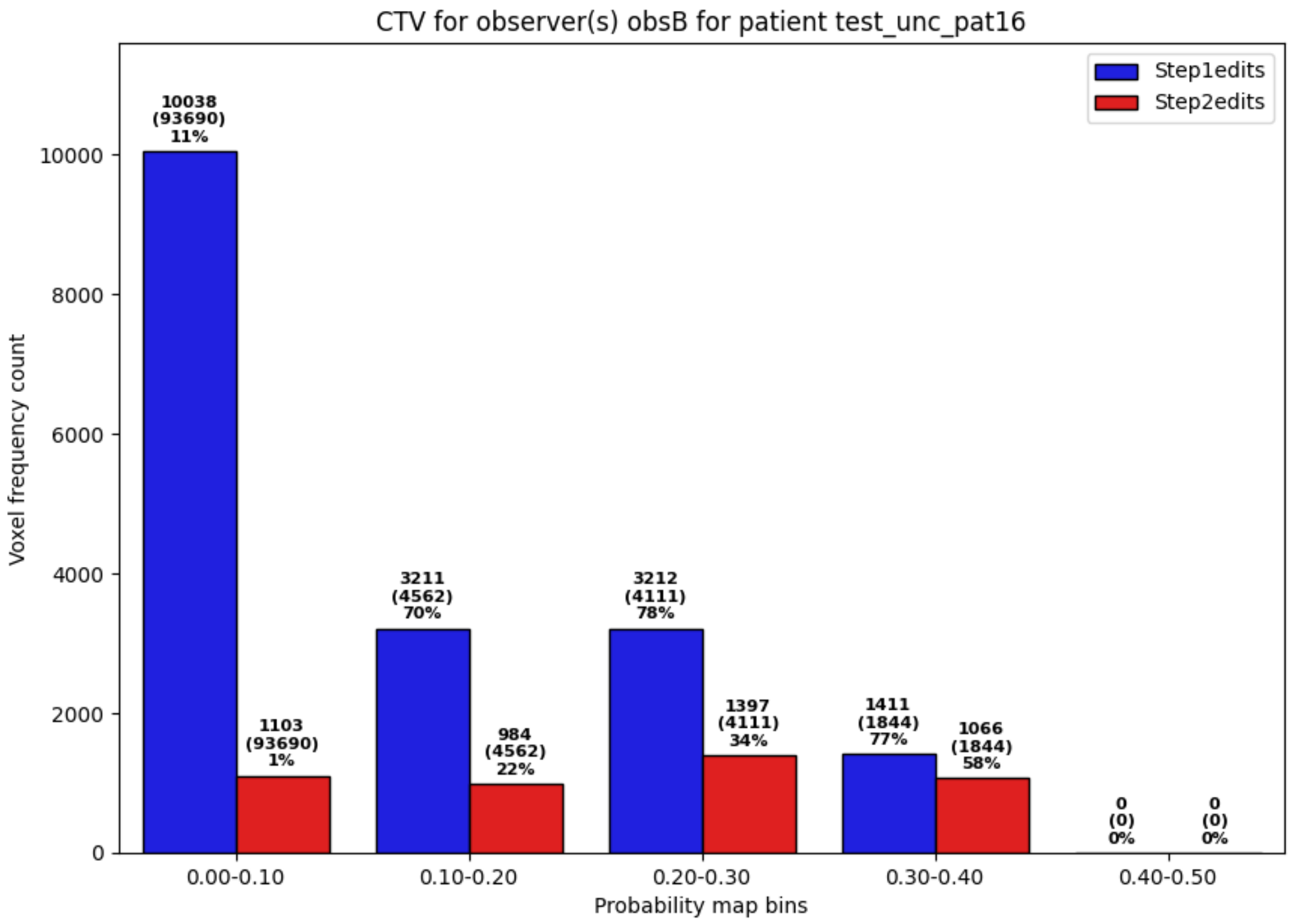

**Fig.S9**. Histogram over changed voxels for the prostate outlier patient. Fewer voxels were changed in step 2. Above each bar is the number of changed voxels, and in parentheses is the total number of voxels existing within that probability range. The ratio is calculated as a percentage.

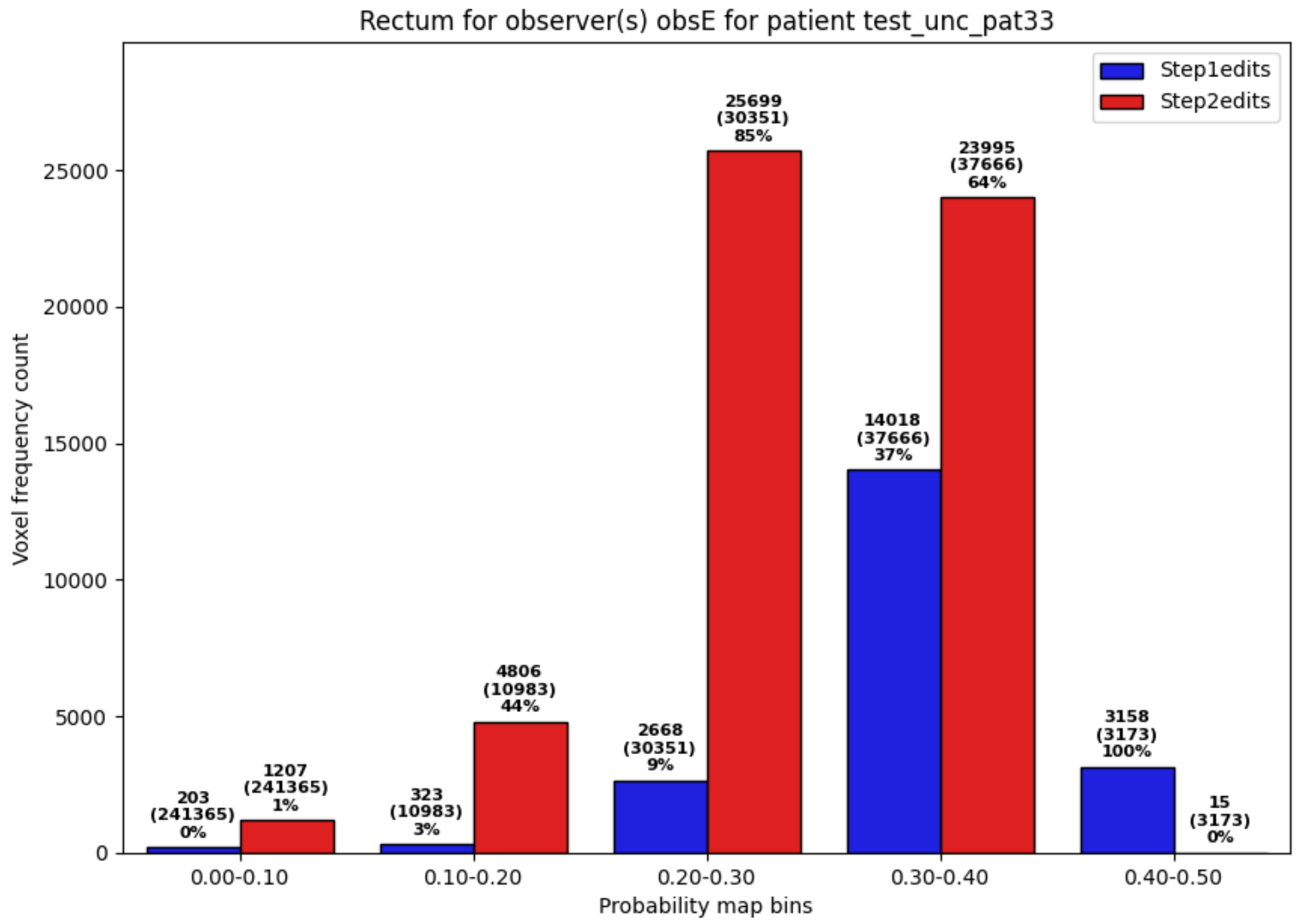

**Fig.S10**. Histogram over changed voxels for the rectum outlier patient. Above each bar is the number of changed voxels, and in parentheses is the total number of voxels existing within that probability range. The ratio is calculated as a percentage.

**Free text answers from oncologists**

The free text answer to the question before step 2 is reported below.

*Do you think delineation uncertainty information will provide you with useful information?*

obsB: ”I am uncertain, but hopefully it will aid in knowing where to look even more carefully.”

obsC: “Yes.”

obsD: “No, I don’t think so.”

obsE: “I think it will help a lot to improve the correctness of my drawing.”

The free text answer to the questions after step 2 is reported below.

*1. Do you think delineation uncertainty information has provided you with useful information?*

obsB: “Yes, on some occasions the "color" enabled me to discover things I had to edit. At other occasions, when I doubled checked an edit I had made with the color, the color made me realize that an edit I had performed was actually not correct.”

obsC: “Very modest.”

obsD: “I don’t think so. There were no surprises in the areas where there were the most uncertainty (base, ventrally in the apex and so on)”

obsE: ”Yes, I found it very useful.”

*2. What was the benefits with delineation uncertainty information?*

obsB: “a) Areas with minimal color can be scrolled through fast. b) I can look more careful at areas with much color. c) I can double-check editing, in particular, if it is outside of the color.”

obsC: “To quickly see where to focus more thoroughly.”

obsD: "I don’t see any. In reality, when drawing (planning), we take into account a lot of information that is missing here, especially tumor characteristics (T stage, histopathology report, location of lesions on MRI and PET with risk for EPE and SVI, where cancer was found in biopsies, etc.) but also patient characteristics (age, comorbidities, erectile function, ...). Delineation uncertainty does not, of course, replace this."

obsE: “I was able to determine the correctness of the drawing faster and with greater certainty.”

*3. What was the drawbacks with delineation uncertainty information?*

obsB: “You feel as if the "allowed" area for editing is only within the color. Areas within the color can easily be edited, but for areas outside the color you need to be very sure you are right before you edit. Unexperienced contourers will rarely/never edit outside the color, which could be problematic it the color is actually wrong.”

obsC: “No obvious but could potentially if incorrect take focus from important areas of interest.”

obsD: “Not really any, except that a few extra seconds were spent looking at it.”

obsE: “I have not experienced any drawback to the information.”

*4. What do you think the potential of delineation uncertainty information is?*

obsB: “To guide your attention towards areas of color/variation = put your time where it is most needed. This benefit has to be balanced against the risk of not editing incorrect delineations that are outside the uncertainty map.”

obsC: “I think they can be used in different ways in quick adaptive workflows.”

obsD: “Unclear. Perhaps of interest to inexperienced doctors for educational purposes. It could be a complement to the clinical information I mentioned in point 2."

obsE: “It can help with daily work.”